\documentclass[aps,twocolumn,superscriptaddress]{revtex4}

\usepackage{amsmath,amssymb,bm}
\usepackage{graphicx}
\usepackage{amsfonts}
\usepackage{color}

\newcommand{\va}[1]{\langle{#1}\rangle}                               

\newcommand{\al}{\alpha}
\newcommand{\be}{\beta}

\begin{document}
\thispagestyle{empty}
 \date{\today}
  \preprint{\hbox{...}}

\title{Comment on  ``Finding the $0^{--}$ Glueball''}

 \author{Alexandr Pimikov}
 \email{pimikov@mail.ru}
 \affiliation{Institute of Modern
 	Physics, Chinese Academy of Science, Lanzhou 730000, China}
 \affiliation{Bogoliubov Laboratory of Theoretical Physics, Joint
 	Institute for Nuclear Research,\\ Dubna, Moscow Region, 141980
 	Russia}

  \author{Hee-Jung Lee}
  \affiliation{Department of Physics Education, Chungbuk National
  	University, Cheongju, Chungbuk 361-763, Korea}

  \author{Nikolai Kochelev}
  \email{kochelev@theor.jinr.ru}
  \affiliation{Institute of Modern Physics,
  	Chinese Academy of Science, Lanzhou 730000, China}
  \affiliation{Bogoliubov Laboratory of Theoretical Physics, Joint
  	Institute for Nuclear Research,\\ Dubna, Moscow Region, 141980
  	Russia}





\thispagestyle{empty}

\noindent \textbf{Comment on ``Finding the $0^{--}$ Glueball''}
\vspace{5mm}

In the Letter~\cite{Qiao:2014vva} the authors explore the mass of
the exotic three-gluon $0^{--}$ glueball within the QCD sum rules (SR) approach~\cite{Shifman:1978bx}.
Even though the Letter presents a relevant discussion on the advantages of this state for experimental observation, the value obtained for the glueball mass is wrong, since the SRs used are inconsistent. Indeed,
combining  Eq.(11) through Eq.(15) of~\cite{Qiao:2014vva} one gets the master SR  for the imaginary part of the correlator
of the glueball interpolating currents in the following form:
\begin{eqnarray}\label{SR1}
\frac{1}{\pi}\int_0^{s_0}\text{Im}\Pi^{\text{(QCD)}}(s)e^{-s\tau}ds
&=&
f_G^2M_0^{12} e^{-\tau M_0^2}\,,
\end{eqnarray}
where $M_0$ is glueball mass, $f_G$ is the decay constant, and $\tau $ is so-called Borel parameter.
The major contribution to the correlator $\Pi^\text{QCD}$ of the A,B,C,D  currents, presented in Eq.(1)-Eq.(4) of ~\cite{Qiao:2014vva},
comes from the Leading Order (LO) part of Operator Product Expansion  for the glueball correlator
 (see Eq(6) and  Eq.(7) in~\cite{Qiao:2014vva}).
 They obtained  the LO part of the correlator  as
\begin{equation}\label{LO}
	\Pi^\text{QCD,LO} = \frac{487}{2^6 3^3 11\cdot 13\pi}\alpha_S^3Q^{12}\ln(Q^2/\mu^2)\,,
\end{equation}
where $\alpha_S$ is the coupling constant, and    $\mu$ is the  momentum scale.
 It is easy to check that the theoretical (left) side in Eq.~(\ref{SR1}) is negative because
 \begin{equation}\label{LO1}
\frac 1\pi \text{Im} \Pi^\text{QCD,LO}(s)= -\frac{487}{2^6 3^3 11\cdot 13\pi}\alpha_S^3s^{6}\,,
\end{equation}
 while the phenomenological (right) part is positive.

Therefore, the SR obtained in~\cite{Qiao:2014vva} is  inconsistent.
We found that the negative sign of the theoretical  part of SR  calculated in~\cite{Qiao:2014vva} is related to the specific  structure of the interpolating currents used. For instance, the  A current has the following form
	\begin{equation} \label{eq:currentv0}
j^A_{0^{--}}(x) \!  = \!  g_s^3 d^{a b c} [g^t_{\alpha \beta} \tilde{G}^a_{\mu \nu}(x)][\partial_\alpha \partial_\beta G^b_{\nu \rho}(x)][G^c_{\rho \mu}(x)]\, ,
	\end{equation}
	where
	\begin{equation}
	g^t_{\alpha \beta}\equiv g_{\al\be}-\partial_\al\partial_\be/\partial^2 .
	\label{projection}
	\end{equation}
	This specific current induces  additional nonphysical poles in the correlator coming
	from the second term of Eq.~(\ref{projection}), which finally leads to the negative imaginary part of the correlator, Eq.~(\ref{LO1}).
Note that the correlator does not depend on the phase of the current
by the definition of the correlators~\cite{Narison:2002pw}:
\begin{equation}\label{cor}
\Pi(Q^2) = i\int\!\! d^4x\, e^{iqx} \va{Tj(x)j^\dagger(0)}\,,
\end{equation}
where Hermitian conjugation $\dagger$ is often omitted considering Hermitian currents.
The above shortcomings lead to the conclusion
	that the currents, presented in~\cite{Qiao:2014vva}, do not couple to gluonic bound states and should not be  considered in
	glueball studies.

Recently a new interpolating current for the exotic three-gluon $0^{--}$ glueball state has been proposed. This current
  leads to consistent SRs and gives a mass for the glueball of 6.3 GeV \cite{Pimikov:2016pag}.

We would like to thank V. Vento for stimulating discussions and useful remarks.
This work has been supported by
Chinese Academy of Sciences President's International Fellowship Initiative
(Grant No. 2013T2J0011 and 2016PM053).
The work by H.J.L. was supported by the Basic Science Research Program
through the National Research Foundation of Korea (NRF) funded by Ministry
of Education under Grants No. 2013R1A1A2009695.
This work was also supported by the Heisenberg--Landau Program (Grant 2016),
the Russian Foundation for Basic Research under Grants No.\ 15-52-04023.

\vspace{.3cm}
\noindent Alexandr Pimikov$^{1,2,*}$, Hee-Jung Lee$^3$ and Nikolai Kochelev$^{1,2,\dag}$

\begingroup\raggedright\leftskip=20pt
{\footnotesize
	\noindent
$^1$Institute of Modern
	Physics, Chinese Academy of Science, Lanzhou 730000, China\\
$^2$Bogoliubov Laboratory of Theoretical Physics, Joint
Institute for Nuclear Research,\\ Dubna, Moscow Region, 141980
Russia\\
$^3$Department of Physics Education, Chungbuk National
	University, Cheongju, Chungbuk 361-763, Korea\\
* pimikov@mail.ru \\
\dag ~kochelev@theor.jinr.ru
}
\par\endgroup

\end{document}